# Anomalous charge transport upon quantum melting of chiral spin order

Y. Fujishiro[1*], C. Terakura[1], A. Miyake[2,7], N. Kanazawa[3], K. Nakazawa[1], N. Ogawa[1], H. Kadobayashi[5], S. Kawaguchi[5], T. Kagayama[6], M. Tokunaga[1,7], Y. Kato[4], Y. Motome[4], K. Shimizu[6], and Y. Tokura[1,4,8*]

[1] *RIKEN Center for Emergent Matter Science (CEMS), Saitama 351-0198, Japan.*
[2] *Institute for Materials Research (IMR), Tohoku University, Ibaraki 311-1313, Japan*
[3] *Institute of Industrial Science, The University of Tokyo, Tokyo 153-8505, Japan*
[4] *Department of Applied Physics, The University of Tokyo, Tokyo 113-8656, Japan*
[5] *Japan Synchrotron Radiation Research Institute (JASRI), Spring-8, Hyogo 679-5198, Japan*
[6] *Center for Quantum Science and Technology under Extreme Conditions, Osaka University, Osaka 560-8531, Japan*
[7] *The Institute for Solid State Physics (ISSP), The University of Tokyo, Chiba 277-8581, Japan*
[8] *Tokyo College, The University of Tokyo, Tokyo 113-8656, Japan*

**A plethora of correlated and exotic metallic states have been identified on the border of itinerant magnetism, where the long-range spin texture is *melted* by tuning the magnetic transition temperature ($T_C$) towards zero, referred to as the quantum phase transition (QPT). So far, the study of QPT in itinerant magnets has mainly focused on low-$T_C$ materials (*i.e.*, typically $T_C \sim 10$ K) where the modification of electronic band structure is subtle, and only makes a small contribution to the QPT. Here we report a distinct example of a magnetic QPT accompanied by a gigantic modification of the electronic structure in FeGe, *i.e.*, a well-studied itinerant chiral magnet hosting near-room-temperature ($T_C$ = 278 K) helical/skyrmion spin texture. The pressure-driven modification of the band structure (*e.g.*, reduction of exchange splitting) is evidenced by magneto-transport study, suggesting a Fermi-surface**

**reconstruction around the magnetic QPT ($P \sim 19$ GPa), in stark contrast to the case of typical metallic ferromagnets. Further application of pressure leads to a metal-to-insulator transition above $P > 30$ GPa, as also corroborated by our density-functional theory (DFT) calculation. Of particular interest is the occurrence of anomalous magneto-transport in the inhomogeneous short-range chiral-spin ground state ($P = 20–30$ GPa) above the QPT, with longitudinal fluctuations of magnetization. The unexpected observation of *spontaneous* anomalous Hall effect in this exotic quantum regime suggests macroscopic time-reversal symmetry (TRS) breaking, even in the absence of long-range magnetic order. Our findings mark the large body of unexplored high-$T_c$ itinerant magnets with broken inversion-symmetry as promising candidates of novel ground state formation near QPT.**

Being known as the origin of ferromagnetism in elemental Fe, Ni, and Co, itinerant magnetism (*i.e.*, splitting of majority/minority spin bands giving rise to magnetism) has been a central focus of study over a century. The research interest on metallic ferromagnets (FMs) was further flourished following the discoveries of correlated and exotic metallic states around QPT (*i.e.*, second order phase transition at zero temperature)[1]. The prime examples include the emergence of non-Fermi liquid (NFL) metallic states (*e.g.*, MnSi, ZrZn$_2$,)[1,2], strange-metal behavior (*e.g.*, CeRh$_6$Ge$_4$)[3], and magnetically mediated unconventional superconductivity (*e.g.*, Uranium-based compounds such as UGe$_2$, URhGe, UCoGe)[4–6]. It is now well established that antiferromagnetic (AF) QPTs are rather ubiquitous[7–9], whereas the genuine QPT for FMs or weak FMs (such as long-period helical magnet) in clean systems is generically avoided, due to the occurrence of first-order phase transition, emergence of antiferromagnetic phase, or spin-glass freezing[1,10]. Instead, the lack of QPT in FMs offers a tantalizing possibility to the formation of new quantum ground states, as shown in the seminal works reporting topological NFL or the partial order state with scalar spin chirality in MnSi above its critical pressure ($P_c \sim 1.5$ GPa)[11–15]. However, the study of QPT in FMs so far has been limited to low-$T_C$ compounds (Extended Data Table 1)[1], partly because exploring QPT in high-$T_C$ compounds required a large $P_c$, which hampers detailed experimental investigations. Thus, a pressure induced QPT on an energy scale large enough to modify exchange band splitting itself (*i.e.*, origin of magnetism in itinerant magnets), has remained largely unexplored, despite the possibility that it could lead to a discovery of a new state of matter.

Here we examine this issue by means of transport measurements of the high-$T_C$ itinerant chiral magnet FeGe under application of pressure using diamond anvil cell (DAC). FeGe belongs to B20-type compounds with cubic chiral crystal structure $P2_13$ (space group #198), where the lattice chirality is characterized by the stacking direction of atoms as viewed from [111] crystal axis (Fig. 1**a**). This chiral crystal structure plays a key role for the emergence of "twisted" spin textures (*e.g.*, magnetic skyrmions and hedgehogs) through the anti-symmetric Dzyaloshinskii-Moriya (DM) interaction[16], as well as the topological chiral fermions which have gained immense attention recently in the field of topological materials science (Fig. 1**b**)[17]. Indeed, FeGe is one of the most-widely studied chiral magnets in the context of magnetic skyrmion formation near room temperature[16,18,19]. the magnetic ground state of FeGe is nearly ferromagnetic, *i.e.* helical spin texture with a long period ($\lambda$ = 70 nm) and a saturated magnetic moment of $m_s$ = 1.0 $\mu_B$/f.u. The skyrmion phase appears in the vicinity of $T_C$, exhibiting the typical magnetic phase diagram of B20-type chiral magnets[18].

We first confirmed that the crystal structure remains intact at least up to $P$ = 49 GPa while the suppression of lattice constant amounts to 6.5 % (Fig. 1**c**, see also Supplementary Section I for experimental details). The variation of lattice constant under pressure shows a striking agreement with our density functional (DFT) calculation. Figure 1**d** shows the calculated electronic structures of FeGe at three different pressure points ($P$ = 0, 35, and 50 GPa), showing a transition from magnetic metal to nonmagnetic insulator at $P \sim 45$ GPa (see Extended Data Fig. 1 for the calculated magnetic moment under pressure). At ambient pressure, FeGe has a metallic band structure with multi-fold fermions located far below the Fermi energy ($E_F$) as highlighted by the red circles. Interestingly, the energy of the multi-fold fermion (*i.e.* three-fold spin-1 fermions under spin-orbit coupling (SOC) and time-reversal symmetry (TRS) breaking)[20–22] is largely enhanced by pressure and tuned to $E_F$ around $P$ = 35 GPa. The system then becomes a non-magnetic narrow gap insulator as shown for the case of $P$ = 50 GPa with an energy gap of ~80 meV. For this paramagnetic state, the multi-fold fermions (*i.e.* six-fold fermions which split into four-fold spin-$\frac{3}{2}$ Rarita-Schwinger-Weyl fermion and two-fold Weyl fermion under SOC) are located above $E_F$ as highlighted by the blue rectangle. The electronic structure here is reminiscent of the isostructural FeSi, which has attracted great interest as a narrow-gap insulator with a non-magnetic spin ground stat[23,24]. Importantly, all these behaviors can be understood in terms of the reduction of exchange splitting (*i.e.* origin of itinerant magnetism) as schematically illustrated in Fig. 1**e.** This transition between magnetic metal and nonmagnetic insulator states in FeSi system has long been studied by chemical doping[25,26] or destructive pulsed–high-magnetic-field experiments ($B$ ~ 500 T)[27]. In this context, our high-pressure experiment provides the first opportunity to study this long-sought metal-insulator transition without introducing any chemical disorder and through detailed magneto-transport measurement. In addition, we discover that magnetic QPT during the metal-insulator transition gives rise to an exotic chiral-spin ground state with broken time-reversal symmetry (TRS) as will be discussed later.

Figure 2**a** shows the temperature dependence of resistivity of FeGe under pressure measured by DAC with a liquid pressure-transmitting medium (see Methods for detail). At $P$ = 1 GPa, FeGe is a good metal with a low residual resistivity ($\rho_0$ ~ 5.7 $\mu\Omega$cm at $T$ = 10 K) and high RRR value (= $\rho_{300\,K}/\rho_{10\,K}$) of about 50 (see Supplementary Section II for the pressure cell dependence). The pressure driven QPT ($P$ ~ 19 GPa), *i.e.*, disappearance of long-range magnetic order (LRMO) revealed by a previous Mössbauer spectroscopy study[28], manifests itself as a rapid enhancement of residual resistivity $\rho_0$ (Fig. 2**b**). Note that the QPT here is not genuine second-order but assumed to be weakly first order[28,29] as in the typical ferroamagnets including MnSi[1,13]. Furthermore, the NFL behavior ($\rho \propto T^{1.5}$) (Supplementary Section III) as well as a broad kink structure in the resistivity below $T$ < 100 K are identified near the QPT, suggesting a possible relation to the short-range magnetic order (SRMO) with an inhomogeneous quasistatic magnetic correlation reported at $P$ = 20–30 GPa[28]. Note that these features of resistivity are consistent with those reported by using Bridgeman-type high-pressure cell[29] as well as our cubic anvil press (Supplementary Section III) measured up to $P$ = 22 GPa. By using

DAC, we successfully extend the pressure regime up to $P \sim 50$ GPa and discover that both the NFL behavior and the broad kink structure are rapidly smeared by the subsequent increase of resistivity towards the nearby insulating state (Fig. 2**a**). The resistivity reaches on the order of $\sim$mΩcm above $P > 30$ GPa in the non-magnetic (NM) phase, being three orders of magnitude larger than that at the ambient pressure. Despite the prominent increase of residual resistivity value upon magnetic transitions (Fig. 2**b**), however, the temperature dependence of resistivity in the NM phase is rather monotonous or quasi-linear, like a lightly doped semiconductor, without showing the activation behavior typically expected for insulating materials (Fig. 2**a**). All these features of zero-field resistivity are summarized in the pressure-temperature phase diagram as shown in Fig. 2**c**. The jump in the resistivity around the QPT ($P = 19$ GPa) as well as the rapid enhancement of the resistivity in the NM phase ($P > 30$ GPa) indicate a clear correlation between the magnetism and electronic structure in this system.

The presence of three magnetic phases (LRMO, SRMO, NM) under pressure is further corroborated by our magneto-transport measurement which has not been reported so far in FeGe under pressure. One clear indication is the emergence of different types of magneto-resistance (MR) in each of the magnetic phases (Fig. 3**a**). In the LRMO phase, kink structure corresponding to helical-FM transition is discernible below $T_C$. The positive MR at low temperature (*e.g.*, $P = 10$ GPa, $T = 10$ K) can be attributed to the cyclotron motion of electron, which gives way to the negative MR due to enhanced spin fluctuations towards $T_C$. In the meanwhile, SRMO phase displays a strong negative MR (*e.g.*, $\sim$23 % at $P = 20$ GPa, $T = 10$ K), with an unusual quasi-linear magnetic-field ($B$) dependence, whose magnitude is the strongest at the lowest temperature. In the NM phase, on the other hand, MR becomes much suppressed and positive, being similar to those reported in FeSi and its doped system[30]. These features are summarized in Extended Data Fig. 2 for more detailed pressure data points. Figure 3**b** shows a contour map of MR in the pressure-temperature phase diagram. It is worthwhile to note that the phase boundary line between LRMO and SRMO shows the largest negative MR, which seems to be continuously connected from $T_C$ at ambient pressure down to the QPT. This result supports the assumption proposed by the Mössbauer spectroscopy study[28]; the chiral magnetic precursor phenomenon around $T_C$ [31,32], *i.e.*, an inhomogeneous chiral-spin state with varying longitudinal modulus of magnetization, is vastly enlarged due to increasing spin fluctuations as FeGe is tuned to its QPT, and eventually forms a ground state at $P$ = 20–30 GPa.

To unravel the nature of this inhomogeneous chiral-spin ground state, *i.e.*, SRMO phase, Hall resistivity measurement was simultaneously performed, which is sensitive to the change in electronic structure as well as the emergent magnetic field due to the scalar spin chirality in spin textures[15]. Firstly, the normal Hall effect (NHE) is significantly altered across the QPT as evidenced by the Hall conductivity ($\sigma_{xy}$), indicating a large modification of the Fermi surface and the strong correlation between the magnetic and electronic phase changes in this system (Fig. 4**a**). For Hall resistivity data, see Supplementary Section IV. It seems that the NHE dramatically evolves throughout the SRMO phase, even leading to a sign change of the dominant carrier (Extended Data Fig. 3). A possible relationship between the observed Lifshitz-like transition at SRMO phase and a metastable electronic structure predicted by our DFT calculation will be discussed in Supplementary Section V. Note that such distinct changes of electronic structure have not been confirmed near QPT in typical 3d-transition metal based itinerant FMs including MnSi[15], suggesting the uniqueness of FeGe. Here, we attribute the *B*-nonlinear behavior of $\sigma_{xy}$ above the QPT (Fig. 4**a**) to the emergence of high-mobility and small electron pocket (see Extended Data Fig. 4 for detail). The estimated mobility $\mu_e$ ~ 750 cm$^2$/Vs ($P$ = 20 GPa and $T$ = 10 K) is a record high value among itinerant magnets, even reaching to that of the ferromagnetic Weyl semimetal Co$_3$Sn$_2$S$_2$[33]. This high-mobility of electron may be attributed to the $E_F$-tuned multi-fold chiral fermion[17,20–22] as discussed in Figs. 1**d,e**.

In addition to this dramatic change of NHE, another striking and unexpected observation would be the spontaneous anomalous Hall effect (AHE) emerging in the SRMO phase (Fig. 4**b**). This suggests a macroscopic breaking of TRS, while the absence of long-range spin texture in this regime is confirmed experimentally[28]. Note that even the long-range helical spin texture in the lower pressure side does not break the macroscopic TRS, *i.e.*, the net magnetization is zero, which explains no spontaneous Hall component in the LRMO phase (Fig. 4**b**). Hence the TRS breaking without long-range magnetic order is highly unusual. Figure 4**c** shows the contour map of the spontaneous AHE component in the pressure-temperature phase diagram. The vicinity of the QPT shows a critical enhancement of the spontaneous QPT, whereas it decays exponentially as pressure (Fig. 4**d**) or temperature (Fig. 4**b**) is tuned away from the QPT. The relationship between longitudinal and Hall conductivity for this AHE does not follow any of those observed in conventional ferromagnets[34]. We also find that the coercive field ($B_c$), corresponding to the hysteresis region, exhibits a critical decay as FeGe is tuned away from the QPT (Fig. 4**e,f**). These results indicate a possibility that the quantum fluctuations are promoting this unconventional spontaneous AHE.

Here we address possible mechanisms for the spontaneous AHE in the SRMO phase. One is the ordered state characterized by a novel order parameter, such as the chiral spin liquid[35] and $Z_2$-vortex order states[36,37]. Indeed, inherent "frustration" in lattice geometry (*i.e.*, trillium lattice with three-dimensional corner-shared equilateral triangle structure[38,39]) as well as in spin interaction in B20-type materials[40] could offer a key role in these intrinsic mechanisms. Another interpretation is that our study offers a metallic counterpart for a rather new concept of "mesoscale pattern formation" near QPT[41]. In particular, the chiral spin interaction combined with a softened amplitude variation of magnetization is predicted to generate an inhomogeneous magnetic phase, which has long been discussed in analogy with the blue phase in chiral nematic liquid crystals[42–44]. However, this skyrmionic ground state alone cannot explain the hysteresis behavior or spontaneous component in AHE, since the droplet-like skyrmion is energetically unstable in a three-dimensional chiral magnet[45]. Indeed, the partial order phase of MnSi exhibit only "*B*-induced" topological Hall effect[15]. In the meanwhile, additional confinement effect due to magnetic domain formation near QPT, for instance, may give rise to spontaneous AHE component. Indeed, mesoscopic spin cluster formation has been confirmed near FM QPT, not only in a strongly disordered systems (*i.e.*, quantum Griffith phase[1]) but also in a stoichiometric compound with minimal chemical/lattice disorder recently[46]. In this context, magnetic inhomogeneity associated with the metastable electronic structure under pressure could serve as an effective disorder, once the effects of electron correlation and quantum fluctuations are considered, which would be beyond the framework of current DFT calculation (Supplementary Section V).

Our study demonstrates QPT with a hitherto unexplored large energy scale, *i.e.*, melting near-room-temperature chiral spin texture by application of pressure in FeGe. In sharp contrast to the previously studied low-$T_C$ itinerant magnets, strong correlation between electronic and magnetic transitions was identified due to significant modification in exchange splitting. This led to (i) realization of a rare pressure-driven "metal-to-insulator" transition, as well as (ii) exotic short-range chiral-spin quantum ground state showing (iii) spontaneous AHE persisting in a wide temperature-pressure region ($T < 100$ K, $P = 20$–30 GPa). We stress that the lack of spatial inversion symmetry, allowing anti-symmetric spin interaction (*i.e.*, DM interaction), should play a key role here when combined with the longitudinal fluctuations of magnetization near QPT[42–44]. Hence our study motivates us to visit the large body of unexplored high-$T_C$ itinerant magnets without spatial inversion symmetry, which may offer quantum phases with new physical properties as demonstrated here. It may also provide a promising route for the realization of exotic states of matter with topological nature, non-Abelian quasiparticle excitations, or unconventional superconductivity at higher temperatures.

## Methods
### Crystal growth
FeGe single crystals used for the high-pressure resistivity experiments were grown by a chemical vapor transport method[47]. Raw ingredients of Fe and Ge powder and transport agent of iodine were heated in an evacuated quartz tube under a thermal gradient between 560 °C and 500 °C for two weeks. FeGe polycrystal used for the high-pressure X-ray diffraction experiment was synthesized by the arc-melting of Fe and Ge chunks, followed by a high-temperature and high-pressure treatment at 800 °C, 5 GPa for 1 hour.

### High-pressure resistivity measurement
The electrical resistivity of single crystalline FeGe at high pressure was measured with the standard four-probe method using a cubic-anvil-type high-pressure apparatus (cell #A) and a diamond anvil cell (DAC) (cell #B,C,D). For the cubic anvil press, we used sintered diamond of $3 \times 3$ mm$^2$ size for the anvil tops. The gold wires were attached to the sample by silver paste as electrodes. The sample was placed in Teflon capsule inside the MoO gasket and Daphne oil 7474 was used as a pressure-transmitting medium. As for the DAC, we used Physical Properties Measurement System (PPMS, Quantum Design) to measure resistivity under magnetic field up to $B$ = 14 T, and the pressure was calibrated by fluorescence of ruby. The diamond anvil with a culet size of $\emptyset 600$ $\mu$m with SUS316L gasket (cell #C) as well as $\emptyset 300$ $\mu$m with rhenium gasket (cell #B,D) were used to generate a maximum pressure of ~18 GPa (cell #C), ~30 GPa (cell #B), and ~50 GPa (cell #D), respectively. Pressure-transmitting medium was Daphne oil 7575, except for cell #B where NaCl was used. The insulation layer was made by cBN powder and epoxy, while Pt or Au foil was used as the electrodes. As for the sample (cell #D) shown in the main text (sample size of ~$18 \times 70 \times 70$ $\mu$m$^3$), we used focused ion beam (FIB) (Hitachi, NB5000) to deposit amorphous Tungsten to make an electrical connection between the foil and sample due to is limited size. By using this original technique, we succeeded in suspending the sample at the starting low pressure regime within the liquid pressure-transmitting medium. As mentioned in Suplementary Section II, we found that using liquid pressure medium was crucial to realize high residual resistivity ratio in FeGe.

**High-pressure synchrotron X-ray diffraction (XRD) experiment**

High-pressure synchrotron XRD experiment up to ~50 GPa was carried out at SPring-8 BL10XU ($\lambda = 0.4136$) in Japan by using DAC with a diamond culet of 300 $\mu$m and a rhenium gasket. Pressure was calibrated by fluorescence of ruby or Raman spectroscopy of diamond. Polycrystalline powder samples were put inside DAC with Helium as a pressure-transmitting medium. The diffraction patterns were collected by a flat-panel detector and integrated to $2\theta$ versus intensity data. Crystal structure refinements were performed by the software RIETAN-FP[48].

**Density functional theory calculation**

The OpenMX code[49,50] was used for the calculations based on the density functional theory (DFT). The Kohn-Sham equation was solved under the generalized gradient approximation proposed by Perdew-Burke-Ernzerhof[51] for the exchange-correlation functional, and norm-conserving and total angular momentum dependent pseudopotentials were employed. We choose pseudo-atomic orbitals Fe5.5H-s3p2d1 and Ge7.0-s3p2d2 as a basis of the Kohn-Sham wavefunctions; see Ref. 49 for the details. To obtain the optimal structure under pressure, we used the quasi-Newton method for the relaxation of the atomic positions and the lattice vectors until the residual force becomes lesser than 0.01 eV/Å per atom within the nonrelativistic calculations. This was performed under two ferromagnetic states (having Fe moment of ~1.2$\mu_B$ and ~0.4$\mu_B$) and the paramagnetic state, and the enthalpy of each state was compared to identify the stable magnetic state at each pressure. The effect of the spin-orbit coupling on the electronic states was considered by performing fully relativistic calculations[52] for the optimal structure and magnetic states under each pressure. The cutoff energy for the fast Fourier transform grid was set to 1200 Ry and the Brillouin zone was sampled with 16×16×16 ***k*** points. We estimated the magnetic moments from the Mulliken population analysis.

## References

[1] Brando, M. Belitz, D., Grosche, F. M. & Kirkpatrick, T. R. Metallic quantum ferromagnets. *Rev. Mod. Phys.* **88**, 025006 (2016).

[2] Stewart, G. R. Non-Fermi-liquid behavior in d- and f-electron metals. *Rev. Mod. Phys.* **73**, 797 (2001)

[3] Shen, B. et al. Strange-metal behavior in a pure ferromagnetic Kondo lattice. *Nat. Phys.* **579**, 51–55 (2020)

[4] Saxena, S. S. et al. Superconductivity on the border of itinerant-electron ferromagnetism in UGe2. *Nature* **406**, 587–592 (2000).

[5] Aoki, D. et al. Coexistence of superconductivity and ferromagnetism in URhGe. *Nature* **413**, 613–616 (2001).

[6] Huy, N. T. et al. Superconductivity on the border of weak itinerant ferromagnetism in UCoGe. *Phys. Rev. Lett.* **99**, 067006 (2007).

[7] Gegenwart, P., Si, Q. & Steglich, F. Quantum criticality in heavy-fermion metals. *Nat. Phys.* **4**, 186–197 (2008).

[8] Daou, R. et al. Linear temperature dependence of resistivity and change in the Fermi surface at the pseudogap critical point of a high-$T$c superconductor. *Nat. Phys*. **5**, 31–34 (2008).

[9] Löhneysen, H. v. et al. Fermi-liquid instabilities at magnetic at magnetic quantum phase transitions. *Rev. Mod. Phys.* **79**, 1015 (2007).

[10] Belitz, D. et al. First order transitions and multicritical points in weak itinerant ferromagnets. *Phys. Rev. Lett.* **82**, 4707 (1999).

[11] Pfleiderer, C., Julian, S. R., Lonzarich, G. G. Non-Fermi-liquid nature of the normal state of itinerant-electron ferromagnets. *Nature* **414**, 427–430 (2001).

[12] Doiron-Leyraud, N. et al. Fermi-liquid breakdown in the paramagnetic phase of a pure metal. *Nature* **425,** 595–599 (2003).

[13] Uemura, Y. J. et al. Phase separation and suppression of critical dynamics at quantum phase transitions of MnSi and $(Sr_{1-x}Ca_x)RuO_3$ *Nat. Phys.* **3**, 29–35 (2007).

[14] Pfleiderer, C. et al. Partial order in the non-Fermi-liquid phase MnSi. *Nature* **427**, 227–231 (2004).

[15] Ritz, R. et al. Formation of a topological non-Fermi liquid in MnSi. *Nature* **497**, 231–234 (2013).

[16] Nagaosa, N. & Tokura, Y. Topological properties and dynamics of magnetic skyrmions. *Nat. Nanotech.* **8**, 899–911 (2013).

[17] Chang, G. et al. Topological quantum properties of chiral cyrstals. *Nat. Mater.* **17**, 978–985 (2018).

[18] Yu, X. Z. Near room-temperature formation of a skyrmion crystal in thin-films of the helimagnet FeGe. *Nat. Mater.* **10**, 106–109 (2011).

[19] Zheng, F. et al. Experimental observation of chiral magnetic bobbers in B20-type FeGe., *Nat. Nanotech.* **13**, 451–455 (2018).

[20] Sanchez, D. S. et al. Topological chiral crystals with helicoid-arc quantum states. *Nature* **567**, 500–505 (2019).

[21] Rao, Z. et al. Observation of unconventional chiral fermions with long Fermi arcs in CoSi. *Nature* **567**, 496–499 (2019).

[22] Takane, D. et al. Observation of chiral fermions with a large topological charge and associated Fermi-Arc Surface States in CoSi. *Phys. Rev. Lett.* **122**, 076402 (2019).

[23] Schlesinger, Z. et al. Unconventional charge gap formation in FeSi. *Phys. Rev. Lett.*

**71**, 1748 (1993).

[24] Tomczak, J. M., Haule, K., Kotliar, G. Signatures of electronic correlations in iron silicide. *Proc. Natl. Acad. Sci. U. S. A.* **109** (9), 3243–3246 (2012).

[25] Anisimov, V. I. et al. First-order transition between a small gap semiconductor and a ferromagnetic metal in the isoelectronic alloy FeSi$_{1-x}$Ge$_x$. *Phys. Rev. Lett.* **89**, 257203 (2002).

[26] Yeo, S. et al. First-order transition from a Kondo insulator to a ferromagnetic metal in single crystalline FeSi$_{1-x}$Ge$_x$. *Phys. Rev. Lett.* **91**, 046401 (2003)

[27] Nakamura, D. Semiconductor-metal transition in correlated narrow-gap semiconductor FeSi driven by an ultra-high magnetic field. *Phys. Rev. Lett.* **127**, 156601 (2021).

[28] Barla, A. et al. Pressure-induced inhomogeneous chiral-spin ground state in FeGe. *Phys. Rev. Lett.* **114**, 016803 (2015).

[29] Pedrazzini, P. et al. Metallic state in cubic FeGe beyond its quantum phase transition. *Phys. Rev. Lett.* **98**, 047204 (2007).

[30] Manyala, M. et al. Magnetoresistance from quantum interference effects in ferromagnets. *Nature* **404**, 581–584 (2000).

[31] Wilhelm, H. et al. Precursor phenomena at the magnetic ordering of the cubic helimagnet FeGe. *Phys. Rev. Lett.* **107**, 127203 (2011).

[32] Pappas, C. Chiral paramagnetic skymrion-like phase in MnSi. *Phys. Rev. Lett.* **102**, 197202 (2007).

[33] Liu, E. et al. Giant anomalous Hall effect in a ferromagnetic kagome-lattice semimetal, *Nat. Phys.* **14**, 1125–1131 (2018).

[34] Nagaosa, N. et al., Anomalous Hall effect. *Rev. Mod. Phys.* **82**, 1539 (2010).

[35] Machida, Y. et al. Time-reversal symmetry breaking and spontaneous Hall effect without magnetic dipole order. *Nature* **463**, 210–213 (2010).

[36] Martin, I., Batista, C. D., Itinerant electron-driven chiral magnetic ordering and spontaneous quantum Hall effect in triangular lattice models. *Phys. Rev. Lett.* **101**, 156402 (2008).

[37] Kawamura, H. Z$_2$-vortex order of frustrated Heisenberg antiferromagnets in two dimensions. *J. Phys.: Conf. Ser.* **320**, 012002 (2011).

[38] Hopkinson, J. M., Kee, H. -Y. Geometric frustration inherent to the trillium lattice, a sublattice of the B20 structure. *Phys. Rev. B* **74**, 224441 (2006).

[39] Isakov, S. V., Hopkinson, J. M., Kee, H. -Y. Fate of partial order on trillium and distorted windmill lattices. *Phys. Rev. B* **74**, 014404 (2008).

[40] Fujishiro, Y., Kanazawa, N., Tokura, Y. Engineering skyrmions and emergent monopoles in topological chiral crystals. *Appl. Phys. Lett.* **116**, 090501 (2020).

[41] Wendl, A. et al. Emergence of mesoscale quantum phase transitions in a ferromagnet. Nature **609**, 65-70 (2022).

[42] Roßler, U. K., Bogdanov, A. N., Pfleiderer, C. Spontaneous skyrmion ground states in magnetic metals, *Nature* **442,** 797–801 (2006).

[43] Tewari, S., Belitz, D. & Kirkpatrick, T. R. Blue quantum fog: chiral condensation in quantum helimagnets. *Phys. Rev. Lett.* **96**, 047207 (2006).

[44] Conduit, G. J., Green, A. G. & Simons, B. D. Inhomogeneous phase formation on the border of itinerant ferromagnetism. *Phys. Rev. Lett.* **103**, 207201 (2009).

[45] Bogdanov, A. N. & Yablonskii, D. A. Thermodynamically stable 'vortices' in magnetically ordered crystals. The mixed state of magnets. *Sov. Phys. JETP* **68**, 101–103 (1989).

[46] Rana, K. et al. Magnetic properties of the itinerant ferromagnet $LaCrGe_3$ under pressure studied by $^{139}La$ NMR. *Phys. Rev. B* **103**, 174426 (2021).

[47] Richardson, M., The partial equilibrium diagram of the Fe-Ge system in the range 40–72 at. % Ge, and the crystallization of some iron germanides by chemical transport reactions, Acta. Chem. Scand. **21**, 2305 (1967)

[48] Izumi, F., Momma, K., Solid State Phenom. **130**, 15 (2007)

[49] Ozaki, T., Variationally optimized atomic orbitals for large-scale electronic Structures, Phys. Rev. B **67**, 155108, (2003).

[50] http://www.openmx-square.org.

[51] Perdew, J. P., Burke, K., and Ernzerhof, M. Generalized Gradient Approximation Made Simple, Phys. Rev. Lett. **77**, 3865 (1996).

[52] Theurich, G. and Hill, Self-consistent treatment of spin-orbit coupling in solids using relativistic fully separable ab initio pseudopotentials, N. A., Phys. Rev. B **64**, 073106 (2001).

## Acknowledgements

We are grateful to Y. Nakamoto and M. Gen for experimental support and M. Mogi, Y. Ohnishi, P. Pedrazzini, H. Ishizuka, S. Nakatsuji, M. Birch, I. Belopolskii, H. Murayama, T. Arima and N. Nagaosa for fruitful discussions. This work was supported by JSPS/MEXT Grant-in-Aid for Scientific Research (Grant No. 22K14011, No. 22K18965, No. 23H04017, No. 23H05431, No. 23H05462, 22H01185, 21K13875), JST FOREST (No. JPMJFR2038), and Mitsubishi Foundation. The synchrotron radiation experiment was performed at the BL10XU of SPring-8 with the approval of the Japan Synchrotron Radiation Research Institute (JASRI) (Proposal No. 2023A1438).

## Author Contributions

The project was conceived by Y.F., C.T., and Y.T.. Single- and poly-crystals of FeGe were synthesized by N.K.. High-pressure resistivity measurement using cubic anvil press was performed by Y.F. and C.T.. High-pressure magneto-transport measurement using DAC was performed by Y. F. and C.T. with significant supports from A.M. and K.S.. Optical characterization required for the DAC experiments were set up by N.O.. High-pressure synchrotron X-ray experiment was done by Y.F., C.T., K.K., S.K., T.K., and K.S.. DFT calculation was performed by K.N., Y.K., and Y.M.. Y.F. wrote the manuscript with supports from A.M., N.K., K.N., T.K., M.T., Y.K., Y.M., K.S., and Y.T.. All the authors discussed the result and commented on the manuscript.

## Competing Interests

The authors declare no competing interests.

## Materials & Correspondence

Correspondence and requests for materials should be addressed to Y.F. and Y.T..

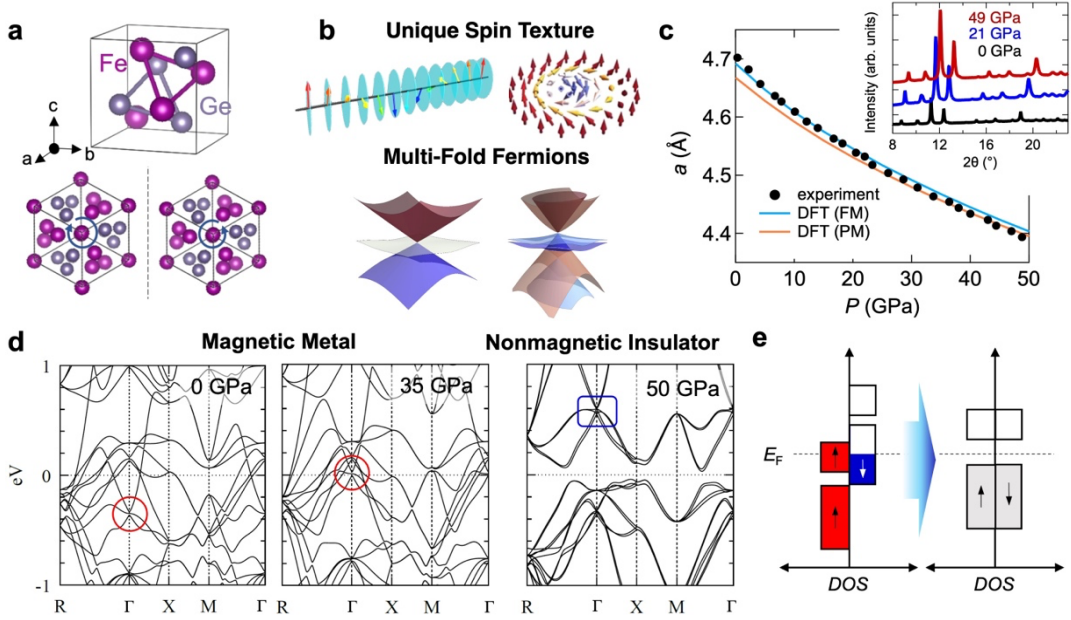

**Figure 1 Crystal structure and electronic structure of FeGe under pressure. a,** Crystallographic unit cell of B20-type FeGe in the cubic chiral space group $P2_13$. The chirality is characterized by the stacking direction of atoms as viewed from [111] direction. **b,** Examples of unique spin textures such (helix and skyrmion) and multi-fold fermions (three-fold (left) and six-fold Weyl fermions which split into four-fold and two-fold fermions (right)) expected in chiral crystals. **c,** Pressure dependence of refined lattice constant in FeGe at room temperature (black dots), showing a striking agreement with those predicted by density functional theory (DFT) calculation for FM state (blue line) at $P < \sim 45$ GPa and PM state (orange line) and $P > \sim 45$ GPa. The inset shows the X-ray diffraction patterns collected at $P = 0, 21, 49$ GPa showing B20-type crystal structure. **d,** Calculated electronic structures at $P = 0, 35,$ and $50$ GPa, exhibiting pressure-induced transition from magnetic metal to nonmagnetic insulator. The red circles represent three-fold spin-1 fermion (under time-reversal symmetry breaking) while the blue rectangle represents six-fold fermions (for the paramagnetic state) which split into four-fold spin-$\frac{3}{2}$ Rarita-Schwinger-Weyl fermion and two-fold Weyl fermions. **e,** Schematic illustration for density of states (DOS) for exchange splitting. At ambient pressure, difference of DOS in majority and minority spins (red and blue regions, respectively) gives rise to magnetism in FeGe, while the reduction of exchange splitting under pressure leads to the non-magnetic state with a gap opening.

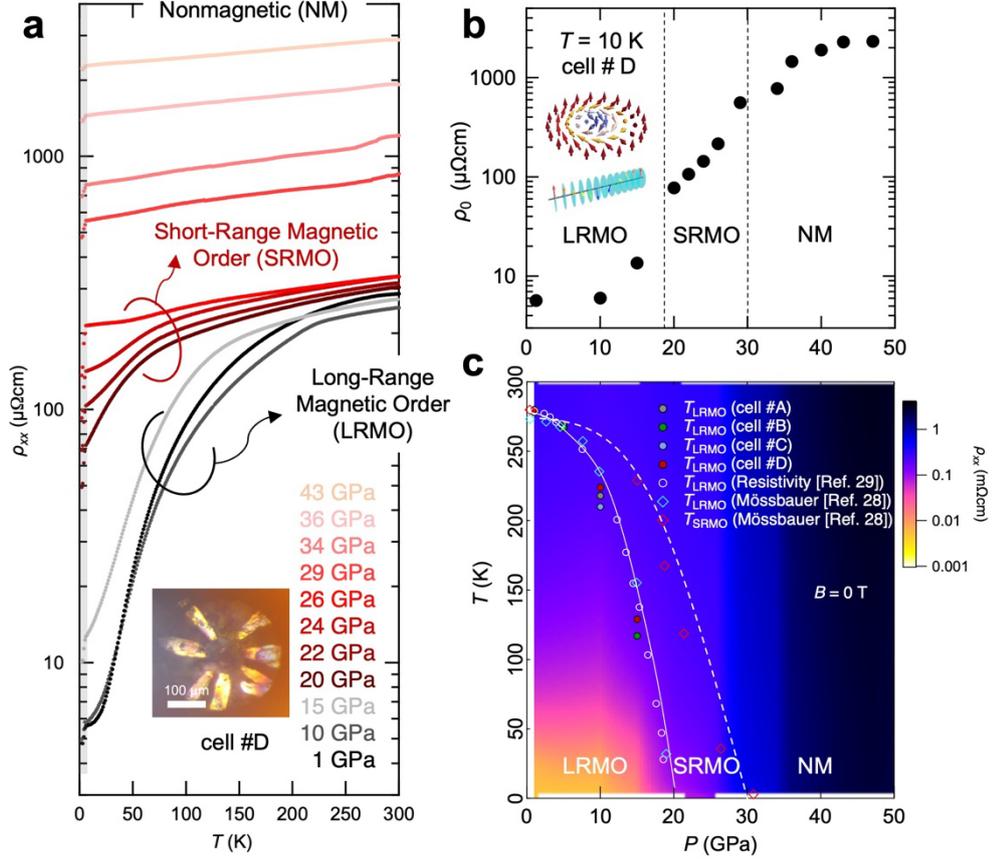

**Figure 2 Temperature dependence of resistivity under pressure. a,** Electrical resistivity measured under pressure by using DAC (cell #D). The inset is a microphotograph of the sample inside DAC with Pt electrodes. The resistivity drop at $T = 7$ K is attributed to the superconductivity of amorphous tungsten used for electrical contacts (see Methods for detail). **b,** Variation of resistivity at $T = 10$ K at various pressures, exhibiting significant increase upon sequential magnetic transitions among long-range magnetic order (LRMO), short-range magnetic order (SRMO), and non-magnetic (NM) phases. The inset shows the schematic illustration of LRMO, *i.e.*, helical spin texture and magnetic skyrmion. **c,** Contour map of resistivity in the temperature-pressure phase diagram. Magnetic transition temperatures ($T_{LRMO}$) for LRMO (helical and skyrmion spin textures) determined from our resistivity measurements as well as those cited from[28,29] are displayed. Inhomogeneous chiral-spin ground state (*i.e.*, SRMO phase) is identified below $T_{SRMO}$ between the LRMO and NM phases[28]. The solid(dashed) line represents the phase boundary between LRMO-SRMO(SRMO-PM) phases.

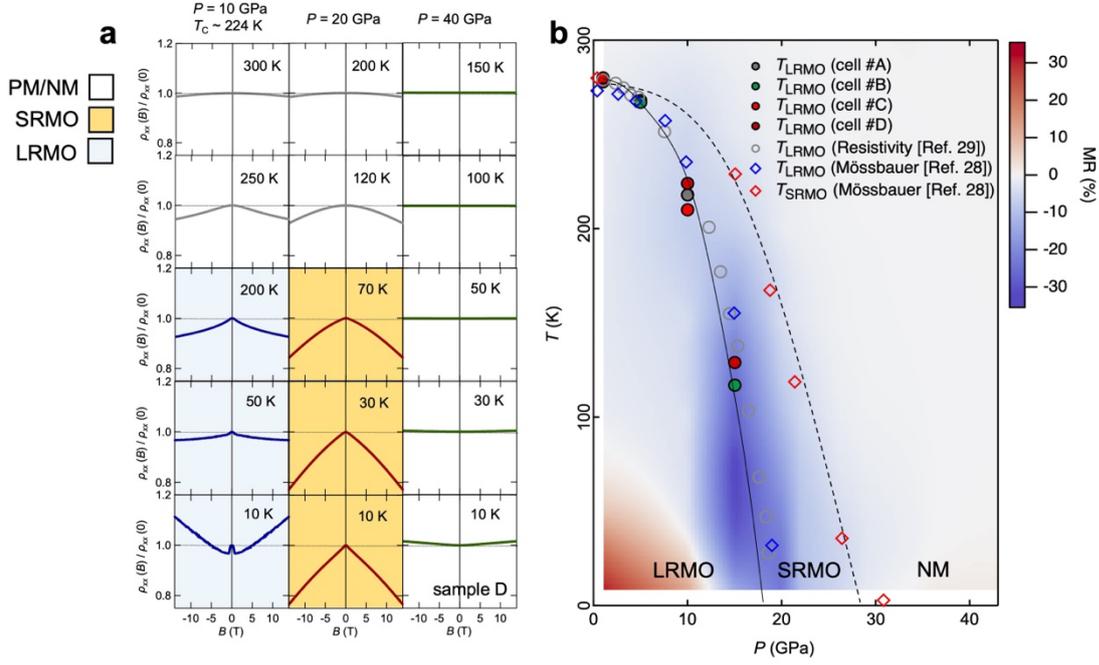

**Figure 3 Magneto-resistance (MR) of FeGe under pressure. a,** MR measured up to $B$ =14 T for $P$ = 10 GPa (LRMO, $T_C$ ~ 224 K), 20 GPa (SRMO), and 40 GPa (NM) phases in cell #D, showing distinct behaviors in terms of sign and $B$-dependence of MR in each of the magnetic phase. The positive MR at low temperatures (e.g. $T$ = 10 K) in LRMO phase is attributed to the cyclotron motion of electrons. Magnetic field is applied perpendicular to the electric current. **b,** Contour map of MR at $B$ = 14 T [*i.e.*, $(\rho_{xx}^{14\,T} - \rho_{xx}^{0\,T})/\rho_{xx}^{0\,T}$] shown in the temperature-pressure phase diagram. Negative MR is critically enhanced along the phase boundary line between LRMO and SRMO, which is maximized near the QPT at $P$ = 19 GPa.

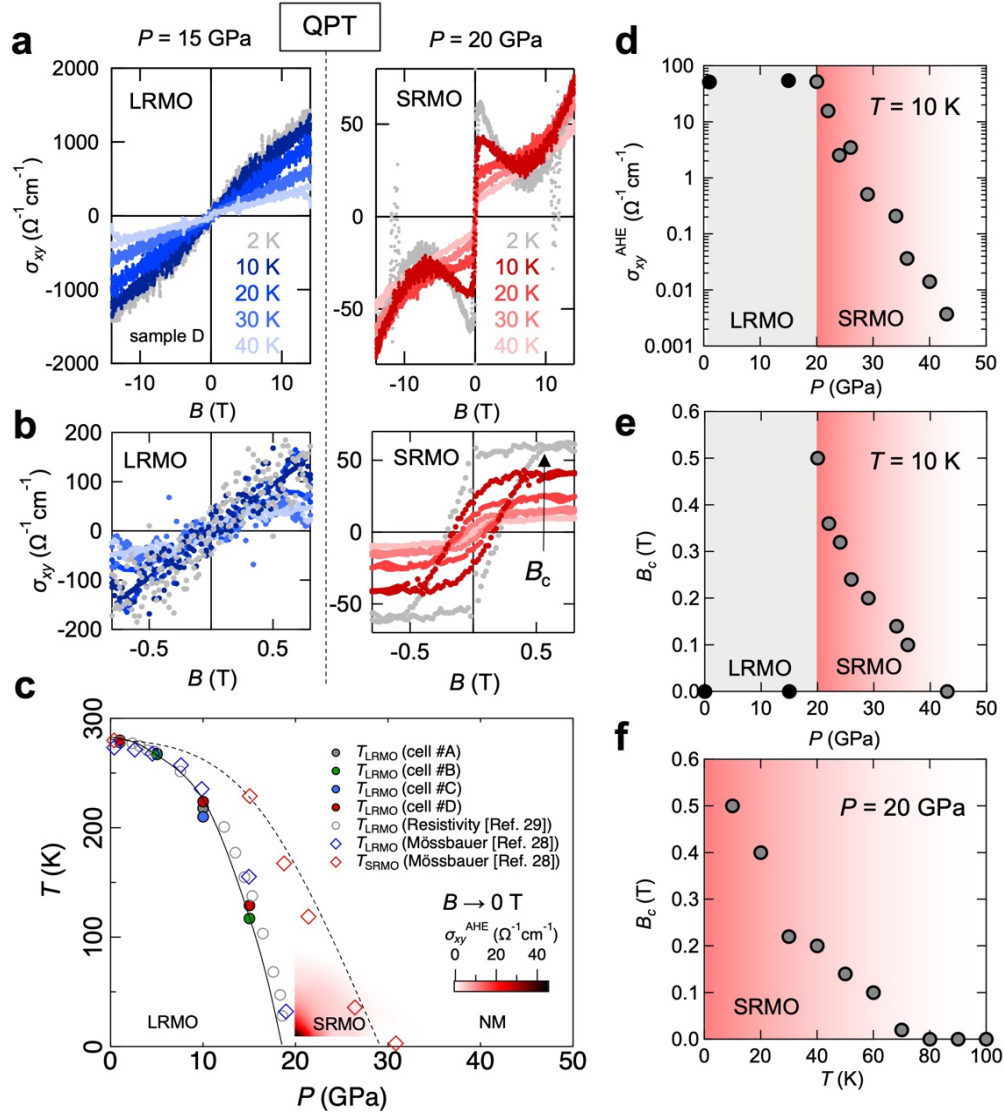

**Figure 4 Hall conductivity of FeGe under pressure. a,** Hall conductivity up to $B = 14$ T for LRMO phase ($P = 15$ GPa) and SRMO phase ($P = 20$ GPa) at various temperatures below $T = 40$ K, showing distinct changes in the magnitude and $B$-dependence across the QPT. **b,** Magnified data near $B = 0$ T for the above panel **a**, where spontaneous component (*i.e.*, anomalous Hall effect (AHE)) with a hysteresis behavior emerges in the SRMO phase. **c,** Contour map of the extrapolated anomalous component of the Hall conductivity obtained by fitting the data above $B > 1$ T. The critical enhancement is discernible near the QPT. **d,** The anomalous AHE at $T = 10$ K as a function of pressure. **e,f,** Coercivity of the hysteresis in AHE showing critical decay as a function of pressure at $T = 10$ K (**e**) and temperature at $P = 20$ GPa (**f**).

| system | space group | $T_C$ (K) | $P_c$ (GPa) | magnetic moment ($\mu_B$) | disorder ($\rho_0/\mu\Omega$cm) | comments |
|---|---|---|---|---|---|---|
| **FeGe** | *P*2$_1$3 (198)* | 278 | 19 | 1.0 | 6 | **NFL, short-range magnetic order, metal-to-insulator transition & spontaneous AHE (this work)** |
| MnSi | | 29.5 | 1.4 | 0.4 | 0.33 | NFL, short-range magnetic order, topological Hall effect in NFL phase |
| ZrZn$_2$ | *Fd*-3*m* (227) | 28.5 | 1.65 | 0.17 | ≥0.31 | NFL |
| Ni$_3$Al | *Pm*-3*m* (221) | 15–41 | 8.1 | 0.075 | 0.84 | NFL |
| CoS$_2$ | *Pa*-3 (205) | 122 | 4.8 | 0.84 | 0.2–0.6 | NFL |
| UGe$_2$ | *Cmcm* (63) | 52 | 1.6 | 1.5 | 0.2 | coexistence of FM+SC |
| UCoGe | *Pnma* (62) | 2.5 | 1.4 | 0.03 | 12 | coexistence of FM+SC |
| U$_3$P$_4$ | *I*4-3*d* (220)* | 138 | 4 | 1.34 | 4 | |
| URhAl | *P*6-2*m* (189) | 25–34 | 5.2 | 0.9 | 65 | |

**Extended Data Table 1 List of stoichiometric ferromagnets showing first-order transitions at pressure-tuned quantum phase transition (QPT).** FeGe studied in this work hosts exceptionally high magnetic transition temperature ($T_C$) and critical pressure ($P_c$) required for achieving QPT. The asterisk symbol* denotes the non-centrosymmetric space group. Many of the compounds listed here exhibit the non-Fermi liquid (NFL) behavior of resistivity near QPT, while some of the uranium-based compounds show coexistence of ferromagnetism (FM) and superconductivity (SC). For other compounds which are tuned by other external parameters such as chemical composition or magnetic field as well as those showing genuine (*i.e.* second-order) QPT, see Ref. 1.

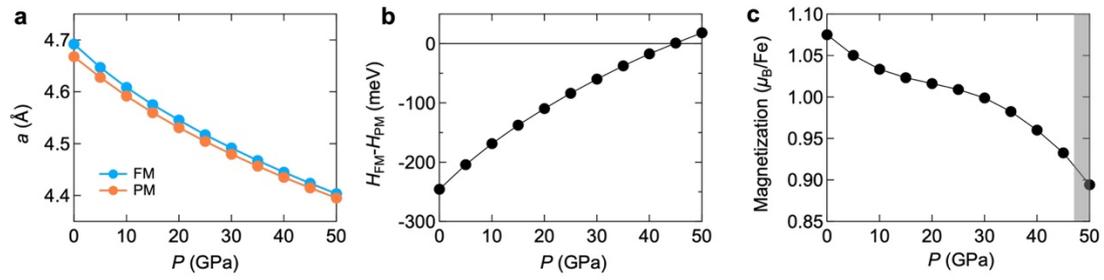

**Extended Data Figure 1 Pressure dependence of calculated lattice constant (a), enthalpy difference (b), and magnetic moment (c) for the ferromagnetic (FM) and paramagnetic (PM) states in FeGe.** The transition from FM to PM state is expected around 45 GPa. The gray shaded area shown in panel (c) represents the expected PM regime.

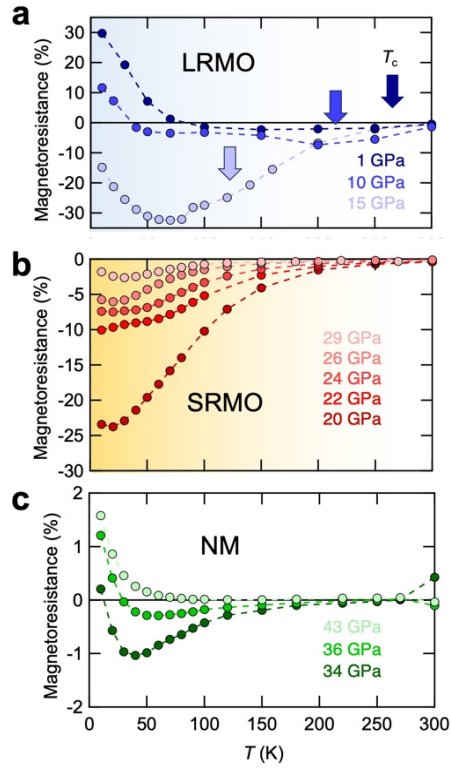

**Extended Data Figure 2 Temperature dependence of magneto-resistance (MR) at $B$ = 14 T at various pressures. a–c**, Detailed data for the temperature dependence of MR at various pressures. For LRMO phase (**a**), the magnetic transition temperature is indicated by the arrow for each pressure, where the negative MR becomes the largest due to enhanced spin fluctuations. The positive MR at low temperatures is attributed to the cyclotron motion of electrons. In the meanwhile, SRMO (**b**) shows larger negative MR as the temperature is lowered. Finally in the NM phase (**c**), the positive MR starts to emerge at low temperatures, although the MR magnitudes are much smaller than in LRMO and SRMO.

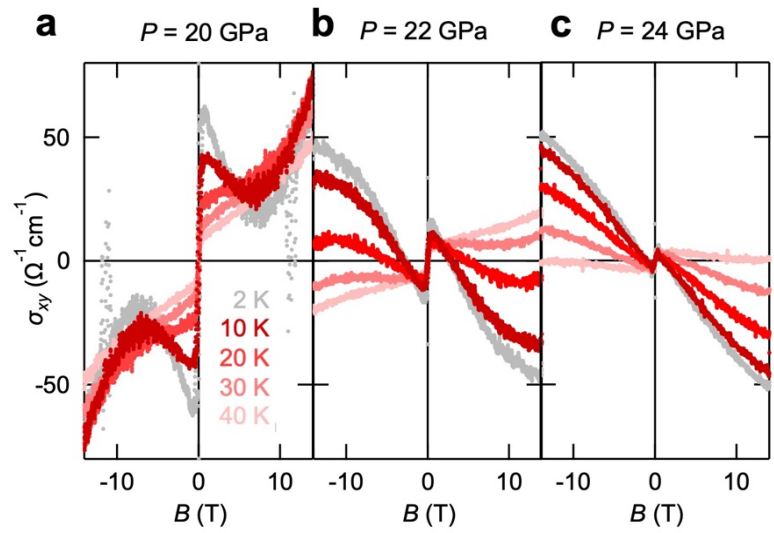

**Extended Data Figure 3 Evolution of Hall conductivity in the short-range magnetic order phase. a–c,** Hall conductivity at $P$ = 20 GPa (**a**), 22 GPa (**b**), 24 GPa (**c**) at various temperatures. The $B$-nonlinear behavior prominent at $P$ = 20 GPa disapepars rapidly as the pressure is increased.

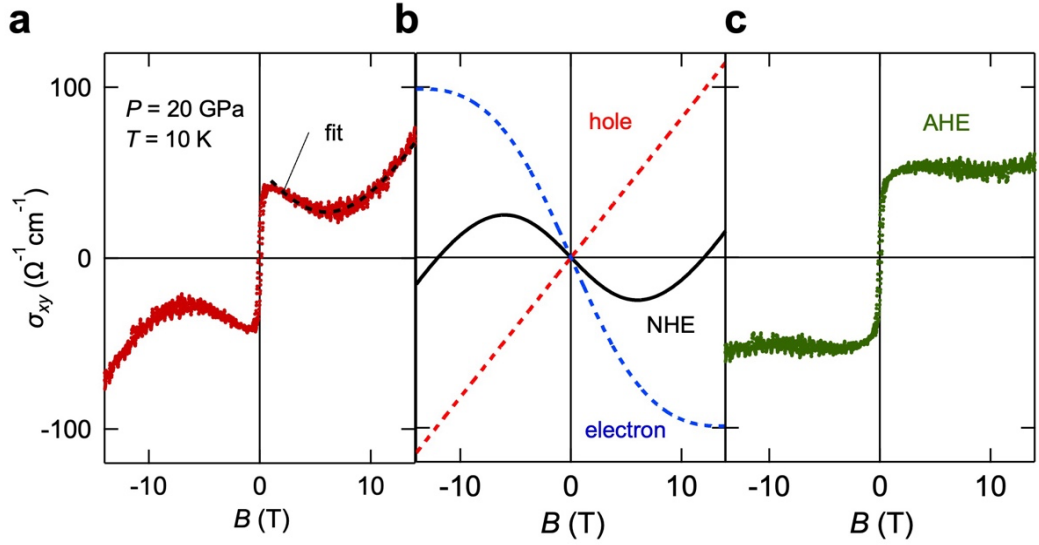

**Extended Data Figure 4 Fitting Hall conductivity by the Drude model. a,** The Hall conductivity at $P = 20$ GPa and $T = 10$ K. The dashed black line is a fitting curve ($B > 1$ T) by the Drude model: $\sigma_{xy} = \frac{\mu^2 neB}{1+\mu^2 B^2} + \alpha B + \beta$. Here, $\mu$ and $n$ represent the mobility and the carrier density, respectively. The latter two terms represent the respective contributions from another (hole-type) normal Hall effect (NHE) and anomalous Hall effect (AHE), *vide infra*. **b,** The extracted normal Hall effect (NHE) component from the fitting is shown in the black line. The blue and red dashed line shows the electron and hole contribution, respectively. The presence of high mobility ($\mu = 750$ cm²/Vs) and low-density ($n = 1.6 \times 10^{18}$ cm⁻³) electron pocket is anticipated from the fitting. **c,** The extracted anomalous Hall effect (AHE) component obtained by subtracting the NHE component (**b**) from the raw data (**a**).